\documentclass[aps,prl,twocolumn,superscriptaddress,showpacs]{revtex4}

\usepackage{graphicx}% Include figure files
\usepackage{dcolumn} % Align table columns on decimal point
\usepackage{bm}      % bold math

\begin{document}

\title{Origin of the phase transition in IrTe$_2$: structural modulation and local bonding instability}

\author{Huibo~Cao}
\email{caoh@ornl.gov}
\author{Bryan~C.~Chakoumakos}
\affiliation{Quantum Condensed Matter Division, Oak Ridge National Laboratory, Oak Ridge, Tennessee 37831, USA}

\author{Xin Chen}
\affiliation{Materials Science and Technology Division, Oak Ridge National Laboratory, Oak Ridge, Tennessee 37831, USA}

\author{Jiaqiang~Yan}
\affiliation{Materials Science and Technology Division, Oak Ridge National Laboratory, Oak Ridge, Tennessee 37831, USA}
\affiliation{Department of Materials Science and Engineering, University of Tennessee, Knoxville, Tennessee 37996, USA}

\author{Michael~A.~McGuire}
\affiliation{Materials Science and Technology Division, Oak Ridge National Laboratory, Oak Ridge, Tennessee 37831, USA}

\author{Hui Yang}
\affiliation{Department of Materials Science and Engineering, University of Tennessee, Knoxville, Tennessee 37996, USA}
\affiliation{Institute of Materials and Metallurgy, Northeastern University, Shenyang, 110004, PR China}

\author{Radu~Custelcean}
\affiliation{Chemical Sciences Division, Oak Ridge National Laboratory, Oak Ridge, Tennessee 37831, USA}

\author{Haidong~Zhou}
\affiliation{Department of Physics and Astronomy, University of Tennessee,
Knoxville, Tennessee 37996-1200, USA}\affiliation{National High Magnetic
Field Laboratory, Florida State University, Tallahassee, Florida
32306-4005, USA}

\author{David J. Singh}
\affiliation{Materials Science and Technology Division, Oak Ridge National Laboratory, Oak Ridge, Tennessee 37831, USA}

\author{David~Mandrus}
\affiliation{Materials Science and Technology Division, Oak Ridge National Laboratory, Oak Ridge, Tennessee 37831, USA}
\affiliation{Department of Materials Science and Engineering, University of Tennessee, Knoxville, Tennessee 37996, USA}

\date{\today}% It is always \today, today,

\begin{abstract}
We used X-ray/neutron diffraction to determine the low temperature (LT) structure of IrTe$_2$. A structural modulation
was observed with a wavevector of \textbf{\textit{k}} =(1/5, 0, 1/5) below $T_s$$\approx$285 K,
accompanied by a structural transition from a trigonal to a triclinic lattice.  We also performed 
the first principles calculations for high temperature (HT) and LT structures, which elucidate the nature of the phase transition and the LT structure. 
A local bonding instability associated with the Te $5p$ states is likely the origin of the structural phase transition in IrTe$_2$. 

\end{abstract}

% Critical phenomena - quantum critical phenomena (superconductivity), 74.40.Kb
% Crystal structure -  effects on transition temperature (superconductivity), 74.62.Bf
% Heavy-fermion solids noncuprate superconductors, 74.70.Tx
% Materials - effects on transition temperature (superconductivity), 74.62.Bf
% Superconducting materials - chalcogenides, 74.70.Xa

\maketitle

The competition between charge density wave (CDW) state and superconductivity is one of mostly interesting phenomena in transition metal dichalcogenides and has been widely studied due to possible relation to high-Tc superconductivity. \cite{CDW1, CDW2, CDW3, CDW4, CuTiSe2} Classically, CDW transitions are second order transitions driven by Fermi surface nesting in a metal, i.e.
a Kohn anomaly leading to a soft mode instability of the high tempearature structure. A key feature of this type of transition is a coupling of the
electronic structure at the Fermi energy to the structural distortion
leading to strong signatures of the phase transition in transport
and also in some systems an interplay between the CDW and
superconductivity.

Recently IrTe$_2$, a new member of the TX$_2$ family incorporating a $5d$ transition metal, presents the superconductivity when its first-order structural transition is suppressed through doping.\cite{PtIrTe,PdIrTe,PhotoIrTe,
PhotoSpecIrTe,PressureIrTe, CuIrTe} Its HT structure has a trigonal symmetry with edge-sharing IrTe$_6$ octahedra forming layers stacked along the $c$-axis with the Ir ions forming an equilateral
triangular lattice (Fig. 1(a)). The LT structure was proposed to be monoclinic based on powder X-ray diffraction.\cite{IrTeStr} Accompanied with the structural transition, the resistivity shows a hump-shaped maximum and the magnetic susceptibility drops, which is
similar to that of the CDW state in other TX$_2$ systems.

However, recent measurements for IrTe$_2$ imply that the physics is more
complicated than a simple CDW.\cite{PtIrTe,PhotoIrTe,PdIrTe,IrTeNMR,PhotoSpecIrTe} In particular, while optical and
transport measurements do imply a strong reconstruction of the
electronic structure at $E_F$ through the transition,
other measurements show that the transition is first order, which is not the generic behavior
of a standard CDW. There are many possible origins for a first 
order transition. One is that the mechanism is still CDW type
related to Fermi surface nesting, but that the transition becomes
first order due to coupling with strain. Another is that it is
driven by local ordering, such as orbital ordering on the
transition metal. Finally, a transition can be driven by chemical
bonding effects. 

Up to now, all the reported studies used a proposed LT structure model from powder X-ray diffraction
\cite{IrTeStr}. Given that electron diffraction revealed the existence of superlattice peaks \cite{PdIrTe}, which principally also can be from the
structure, the LT structure is probably more complicated than the proposed model. The correct LT structure of IrTe$_2$ is essential to explore the
origin of the structural transition and answer its relationship with the superconductivity induced by doping. 

In this work, we report the LT structure of IrTe$_2$ determined by using both single crystal neutron/X-ray and
powder X-ray diffraction. A modulated LT structure was solved and is distinctly different from the
widely used structure model\cite{IrTeStr}, where the superstructure was not observed. We also performed the first principles calculations for the HT and LT structures, which elucidate the nature of the phase transition and the LT structure.

\begin{figure}
\linespread{1}
\par
\begin{center}
\includegraphics[width=3.4 in]{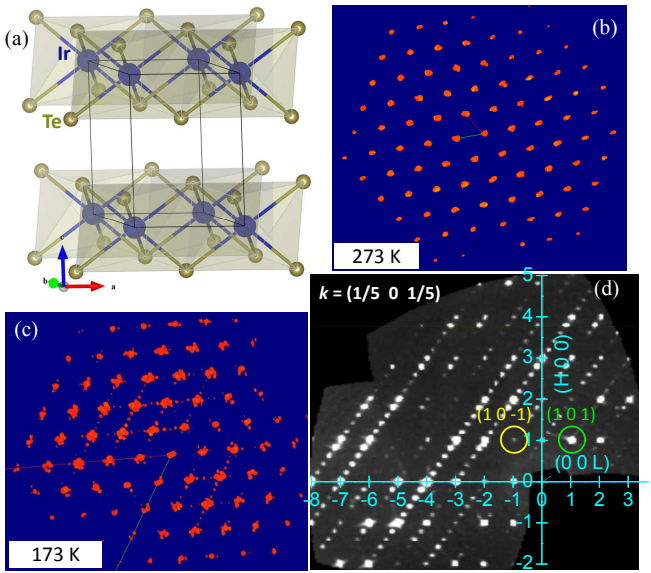}
\end{center}
\par
\caption{(Color Online) (a) The lattice structure of IrTe$_2$ in the trigonal phase at $T$ $>$ $T_s$, brown/blue balls represent Te/Ir atoms. (b) Single crystal X-ray diffraction pattern at 273 K and (c) at 173 K. (d) (H 0 L) reciprocal plane from the pattern in (c) shows the superlattice peaks along the wavevector of (1/5, 0, 1/5). Subcell peaks (1 0 1) and (1 0 -1) are circled in green and yellow to distinguish (H 0 L) from (H 0 -L).}
\end{figure}

\begin{figure}
\linespread{1}
\par
\begin{center}
\includegraphics[width=3.4 in]{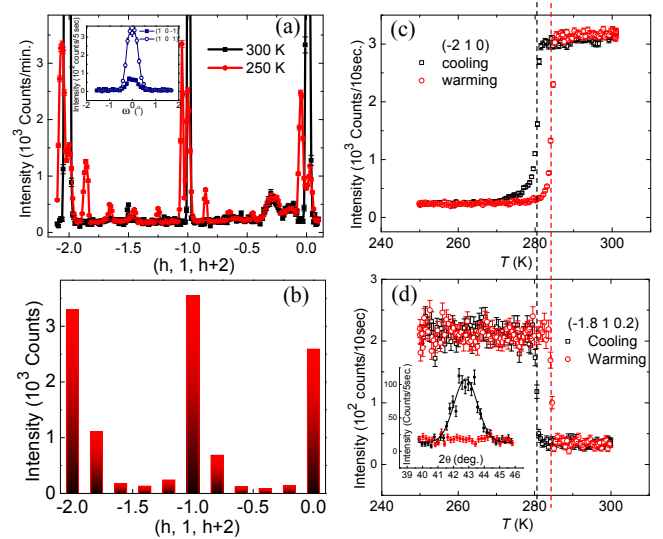}
\end{center}
\par
\caption{(Color Online) (a) \textbf{\textit{q}}-scans along (H 0 H) measured by neutron diffraction at 250 K (black solid square) and 300 K (red solid circle). The scan at 250 K shows the same wavevector of (1/5, 0, 1/5). The strong peak (1 0 1) versus the weak peak (1 0 -1) are also shown in the inset. (b) \textbf{\textit{q}}-scans along (H 0 H) from neutron diffraction calculation based on the solved LT structure (see Fig. 3(b)). (c) and (d)Peak intensities of (-2 1 0) and (-1.8 1 0.2) with temperature warming (red open circle) and cooling (black open square) show the first order structural transition accompanied by the superlattice structural modulation. Black and red dashed lines are guides to the eye. Inset in (d) shows radial scan of (-1.8 1 0.2) at 300 K (red solid circle) and 250 K (black solid square).}
\end{figure}

Single crystals of IrTe$_2$ were grown using self-flux methods as reported \cite{PhotoSpecIrTe}. The polycrystalline sample was prepared by standard solid-state reaction. Single-crystal neutron diffraction was performed at the HB-3A four-circle diffractometer at the High Flux Isotope Reactor at Oak Ridge National Laboratory.
A neutron wavelength of 1.542~\AA~ was used from a bent perfect Si-220 monochromator \cite{hb3a}.
Single-crystal X-ray diffraction was performed using a Bruker SMART APEX CCD diffractometer with Mo K$\alpha$ radiation.
To solve the LT structure, a powder sample was measured with a PANalytical X'Pert MPD diffractometer with an incident beam monochromator (Cu K$\alpha_1$) and an Oxford Phenix Cryostat. Data were collected at 50 and 300 K. The program
Jana2006 was used to solve and refine the LT structure \cite{Jana2006}. The first principles calculations were performed within density functional
theory using the generalized gradient approximation of Perdew, Burke and Ernzerhof (PBE). \cite{pbe}
We used the general potential linearized augmented planewave method \cite{singh-book}
as implemented in the WIEN2k code \cite{wien2k} for calculations of the electronic structure and for structure relaxation. The VASP code
\cite{vasp2}
with projector augmented wave (PAW) pseudopotentials \cite{paw}
with an energy cutoff of 300 eV was used
for the phonon calculations.
The phonons were obtained using a supercell approach \cite{parlinski}
as implemented in the PHONOPY code \cite{togo} with 
a 4x4x3 supercell.

The structural transition in the crystal used for X-ray diffraction occurs near $T_s$ $\approx$ 264 K during the warming process. 
Above $T_s$, trigonal $P\bar{3}m1$ symmetry was observed
(Fig. 1(b)) by X-rays. Below $T_s$, at $T$=173 K, superlattice peaks appear and the subcell peaks split into more than 4 overlapped peaks
(Fig. 1(c)), which makes it difficult to determine the LT lattice and to extract the peak intensity for each \textbf{\textit{q}}. By making a reciprocal plane cut,
the superlattice peaks with a wavevector of (1/5, 0, 1/5) were observed (See Fig. 1(d)). The reflections (1 0 -1) and (1 0 1) have the same \textbf{\textit{q}} length
but different intensities, which are marked in the plot and were used to determine the wavevector of (1/5, 0, 1/5). Note, the wavevector of (1/5 0 -1/5), reported by electron diffraction \cite{PdIrTe}, is equivalent to (1/5, 0, 1/5) by using different Te coordinates.

\begin{figure}[tbp]
\linespread{1}
\par
\begin{center}
\includegraphics[width=3.4 in]{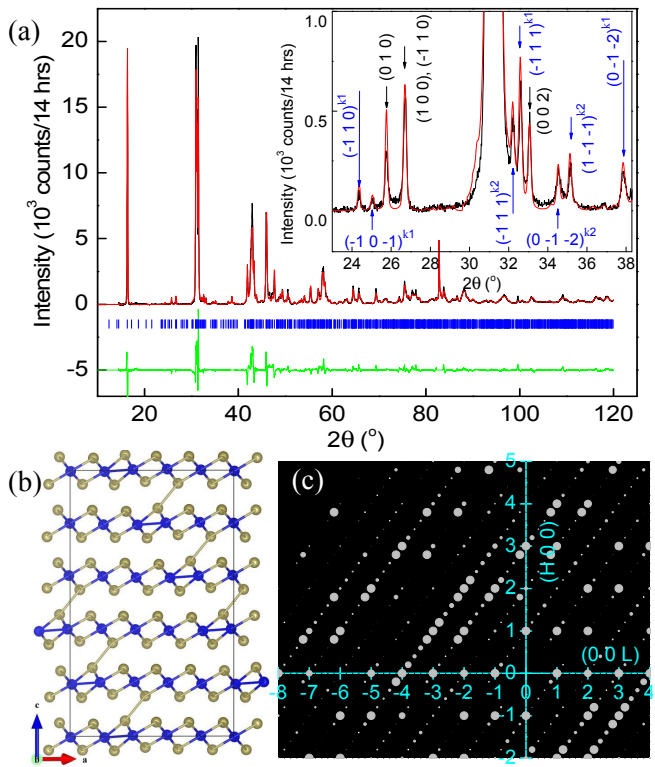}
\end{center}
\par
\caption{(Color Online) (a) Refined powder X-ray patterns at 50 K by the program Jana2006. Black and red patterns are observed and calculated, respectively. The difference is plotted in green. Blue lines mark the Bragg peak positions. The inset shows a specific data region ($2\theta$ in 22$^\circ$-38$^\circ$) highlighting the subcell and satellite Bragg peaks, indicated by black and blue arrows, respectively. (b) the LT superstructure. Brown/blue balls represent Te/Ir atoms, the shortest Te-Te bonds and Ir-Ir bonds are shown by brown and blue lines, respectively. Ir-Te bonds are shown in brown-blue bicolor lines. (c) (H 0 L) reciprocal plane of the calculated X-ray pattern based on the LT structure.}
\end{figure}

To determine the lattice parameters of the LT structure and also confirm the structural origin of the superlattice peaks, we selected a crystal with
a size of $2.1 \times 1.6 \times 0.24$ ~mm for the single crystal neutron diffraction. The \textbf{\textit{q}} scans along (H 0 H) at 300 K
and 250 K are plotted in Fig. 2(a), the superlattice peaks occur below $T_s$ with the same wavevector of (1/5, 0, 1/5), which indicates
that the superlattice peaks found by electron diffraction studies \cite{PdIrTe} and our X-ray diffraction are mainly caused by
the structural modulation. Figure 2(a) inset shows the reflections of (1 0 1) and (1 0 -1) to confirm that the reciprocal lattice is indexed
in the same way as that used for the X-ray data. The structural transition was tracked by both the subcell peak (-2 1 0) and the superlattice peak (-1.8 1 0.2).
The superlattice peaks occur together with the structural transition. Comparing with the crystal measured by X-rays, a higher transition
temperature occurs at $T_s$=285 K during the warming process (Fig. 2(c),(d)). Since the structure refinements above $T_s$ show the same trigonal 
structure, the higher $T_s$ is likely to be associated with the residual stress caused by a faster cooling rate during the sample synthesis. Both crystals should have the same LT structure since they show almost the same diffraction pattern below $T_s$. The hysteresis with warming and cooling confirms that
the structural transition is first-order. Below $T_s$, by analyzing the Bragg peaks measured in the single crystal neutron diffraction, 
we obtained the trial lattice parameters of the LT structure and they were used for indexing the powder diffraction pattern. 
Due to complicated twinning (more than four domains) in the single crystal below $T_s$, the structure factors have not been successfully extracted.

Since the powder X-ray diffraction is insensitive to twinning, we collected data from a powder sample above and below $T_s$. At 300 K, the powder pattern was
collected in an hour and the refinement shows the sample has trigonal symmetry, same as the above measurements. The lattice parameters are \emph{a}=3.9293(1)~\AA, 
 \emph{c}=5.3981(1)~\AA. Supplemental material contains the refined powder pattern and the detailed structural information.  
 A high quality powder diffraction pattern below $T_s$ was 
 measured at 50 K for 14 hrs. The refined pattern is shown in Fig. 3(a). With the lattice parameters and the wavevector obtained from the single crystal 
 diffraction, we were able to index the pattern, index labels are in the inset. Besides the subcell Bragg peaks, the first order
 satellite Bragg peaks (\emph{$k_1$}= (1/5, 0, 1/5)) and the second order satellite Bragg peaks (\emph{$k_2$}= (2/5, 0, 2/5)) are both indicated. The program Jana2006 was
  used to solve and refine the commensurately modulated LT structure in (3+1) dimensional space \cite{Jana2006}. The $R$-factors of the refinement are $R$=0.036 (for the total), $R_0$=0.034 (for the subcell peaks only), $R_1$=0.037 (for the first order satellite peaks), and $R_2$= 0.039 (for the second order satellite peaks), respectively. The
   goodness of the refinement is also shown by the difference of the observed and calculated pattern in Fig. 3(a). With the refined superstructure (Fig. 3(b)), we calculated the single crystal neutron and X-ray diffraction patterns. The \textbf{\textit{q}} scan along (H 0 H) by neutrons and the [H 0 L] reciprocal plane cut by X-rays are plotted in Fig. 2(b) and Fig. 3(c), respectively. They reproduce well the observed patterns in Fig. 2(a) and Fig. 1(d). The LT structure has a triclinic lattice in $P1$ symmetry and its superstructure lattice parameters are refined as \emph{a$_s$}=19.063(3)~\AA, \emph{b$_s$}=3.9545(5)~\AA, \emph{c$_s$}=27.089(3)~\AA, $\alpha_s$=88.74(2)$^\circ$, $\beta_s$=90.49(2)$^\circ$, and $\gamma_s$= 118.99(2)$^\circ$. The volume of the subcell unit below $T_s$ is significantly smaller than the one above $T_s$  by 1.2$\%$, which explains the pressure effects in our earlier work \cite{PressureIrTe}. Pressure favors the phase with a smaller cell volume and so the structural
transition temperature was increased by applying pressure. While the widely used monoclinic LT structure model\cite{IrTeStr} has the similar cell volume as the HT structure, which cannot explain the pressure experiment. Also in the solved LT structure (Fig. 3(b)), one Ir-Ir 'bond'out of 5 and a part of interlayer Te-Te bonds are shortened, which is distinctly different from the previously proposed structure and is important for understanding the true origin of the structural transition in IrTe$_2$.

With the above solved structures, we perform the first principles calculations for the electronic structure and structure relaxation.
We well converged basis sets consisting of standard
LAPW functions up to a cutoff
$RK_{max}$=9, where $K_{max}$ is the planewave sector cutoff and
$R$ is the minimum LAPW sphere radius; the sphere radii were
2.25 Bohr for Ir and 2.50 Bohr for Te. Additionally, local orbitals
were added for the Ir $5p$ and Te $4d$ semicore states.\cite{singh-lo}

\begin{figure}
\includegraphics[width=\columnwidth]{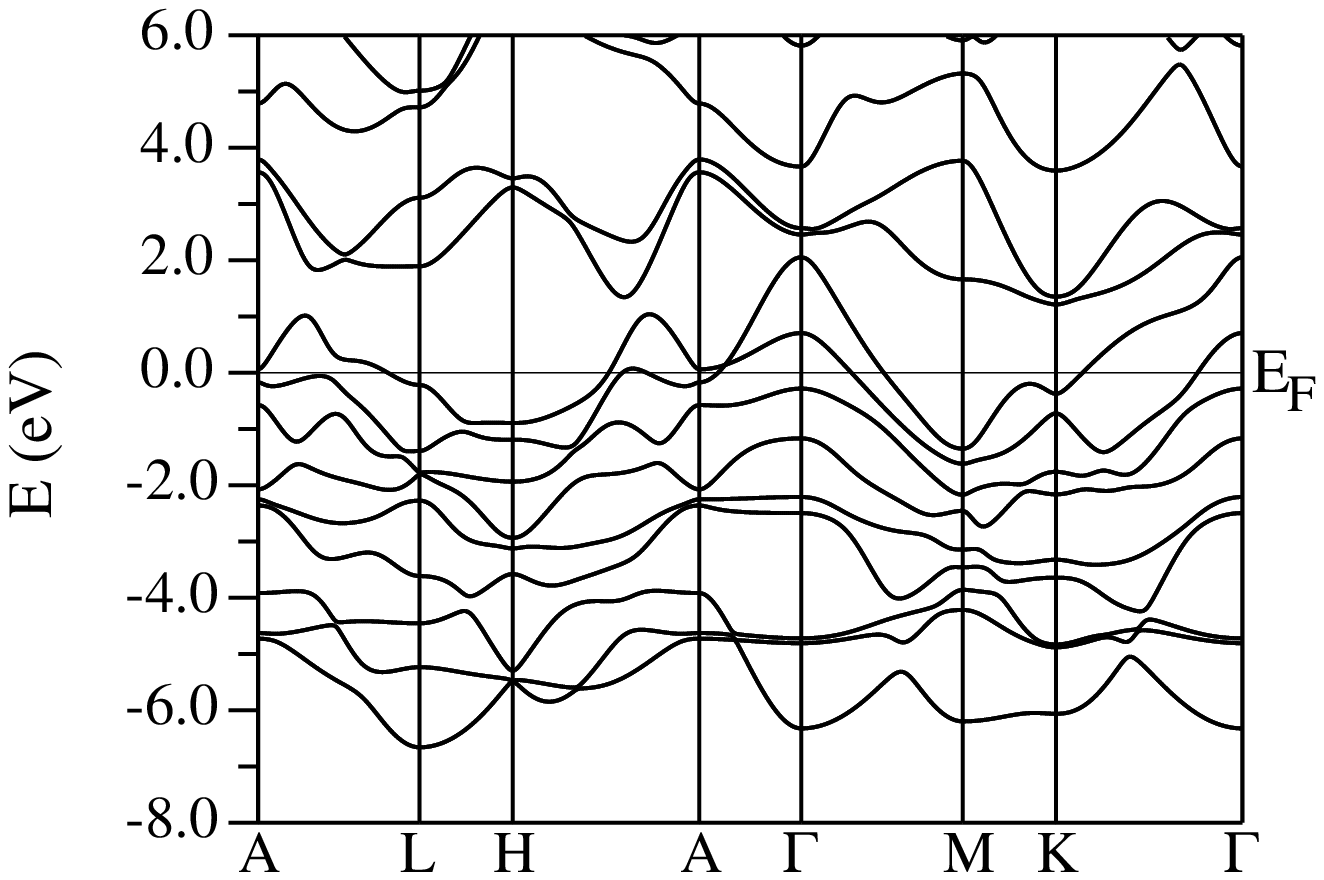}
\includegraphics[height=\columnwidth,angle=270]{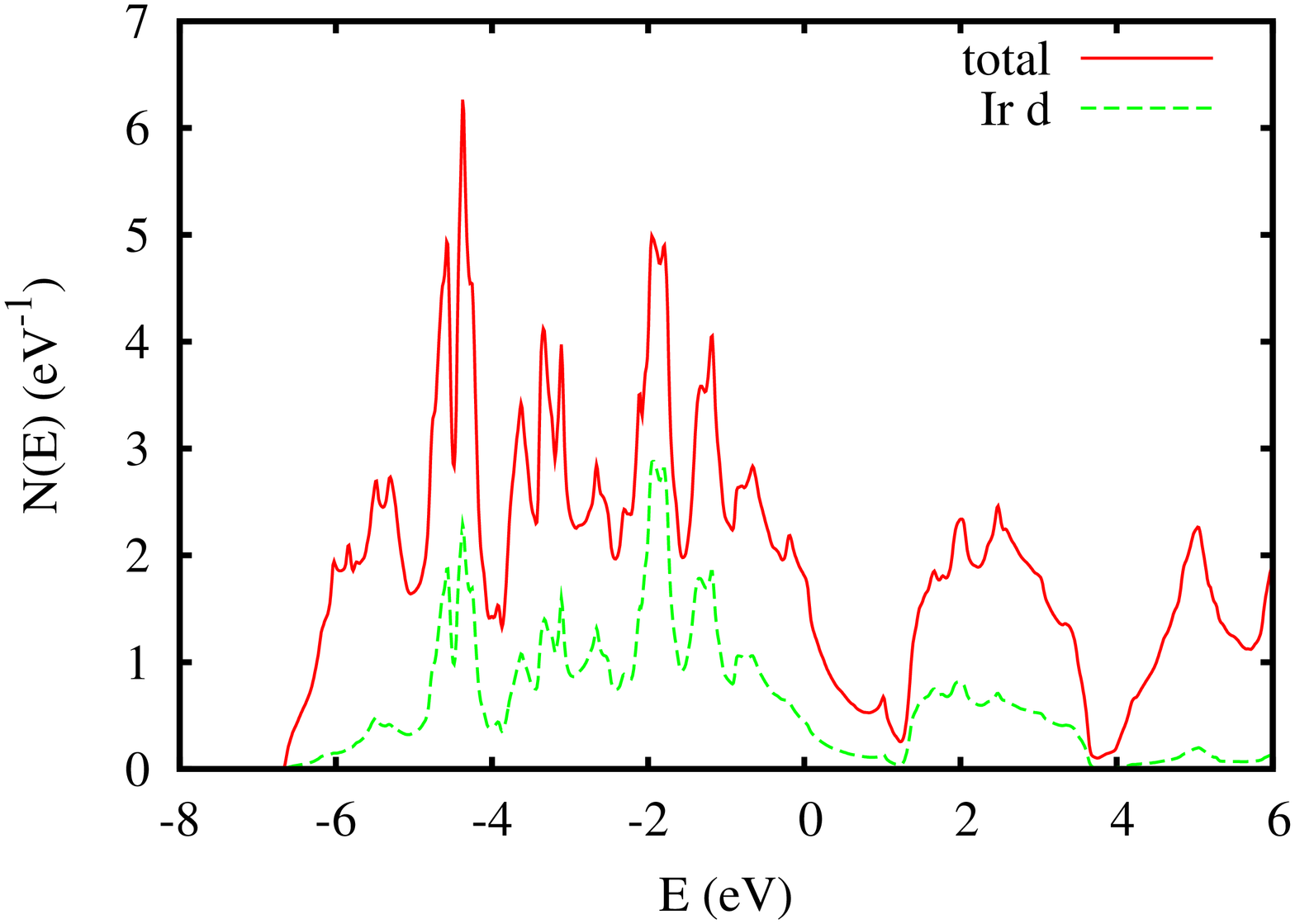}
\caption{(color online) Calculated band structure of IrTe$_2$
HT trigonal structure. Spin orbit is included. The lower panel shows 
electronic density of states for the
HT phase along with the Ir $d$ projection onto the
LAPW sphere. }
\label{bands}
\end{figure}

The band structure for the HT trigonal phase, including
spin orbit, is shown in Fig. \ref{bands}. It is similar
to the previous report of Fang and
co-workers. \cite{PhotoSpecIrTe}
As may be seen, several bands
cross the Fermi energy, $E_F$.
The corresponding electronic density of states and Ir $d$ projection
onto the LAPW sphere is shown in Fig. \ref{bands}.
An examination of the projections of the DOS reveals that
the bands are of hybridized Ir $5d$ - Te $5p$ character over most of the
valence band region shown,
specifically
strong hybridization was
noted early on by Jobic and co-workers. \cite{jobic}

The consequence of this strong
hybridization is that
there is no clear separation into nominally Ir derived and nominally
Te derived bands. Therefore one should not apply nominal ionic models,
e.g. Ir$^{4+}$Te$^{2-}_2$, to understand this compound.
This is reflected in the bond valence sums (see Supplemental material), \cite{brown}
which deviate strongly from the nominal ionic values of 4 and 2 for
Ir and Te, respectively.
This is not surprising both because as a 5$d$ element Ir may be
expected to show substantial hybridization with ligands due to
its extended $d$ shell, and also because as a late transition
element it has a high electronegativity of 2.20 (Pauling scale),
which is comparable to and actually larger than the value of 2.1 for
metalloid element, Te.
Importantly, this means that one has partially filled Te 5$p$
shells in this compound and as a result Te-Te bonding is expected to
be important.

Importantly, the bands around $E_F$ show dispersion both
in the plane of the IrTe$_2$ layers, but also rather strongly in the
$c$-axis direction. This is seen in the dispersions along the
$\Gamma$-$A$ line and also in the differences in dispersions along
the $A$-$L$-$H$-$A$ path from those along
$\Gamma$-$M$-$K$-$\Gamma$. This reflects bonding between the Te atoms in
neighboring layers. This bonding is also clearly reflected in the
crystal structure. The nearest Te-Te distance is 3.49 \AA, and is
across the gap between adjacent IrTe$_2$ sheets. The nearest distance between
Te atoms across a sheet is 3.55 \AA, while that within a hexagonal Te layer
on one side of a IrTe$_2$ sheet is 3.93 \AA, i.e. much longer.
In any case, such three dimensionality of the
electronic structure is not favorable
for nesting and distinguishes the present compound from CDW materials
such as NbSe$_2$ where some bands crossing $E_F$ are three dimensional
but the bands comprising the nested Fermi surface are rather two dimensional.
\cite{johannes,inosov}

\begin{figure}
\includegraphics[height=0.95\columnwidth,angle=270]{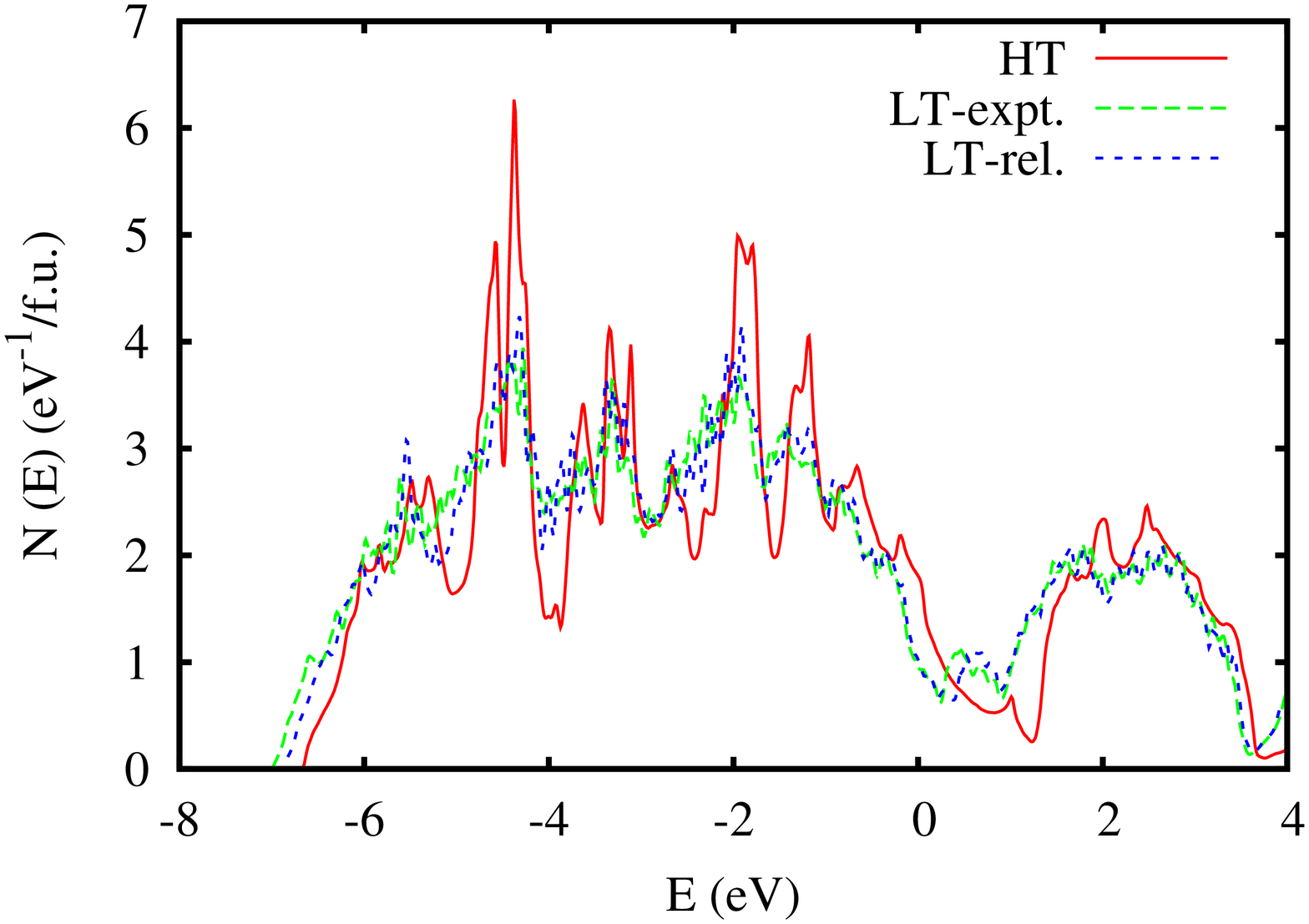}
\includegraphics[height=\columnwidth,angle=270]{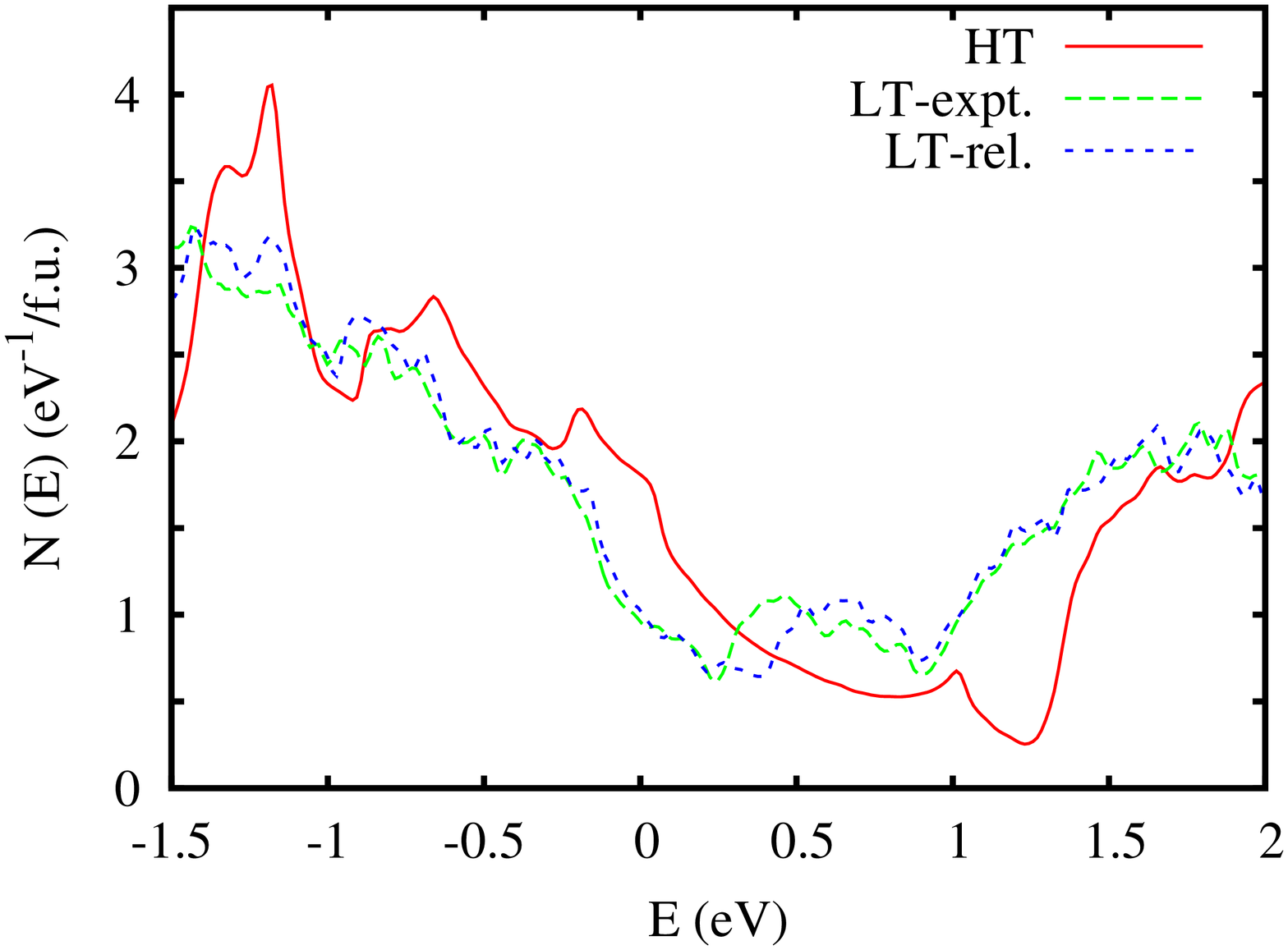}
\caption{(color online) Calculated electronic density of states for
IrTe$_2$ in the HT structure, the LT structure.
The lower panel is a blow up near $E_F$.}
\label{dos-comp}
\end{figure}

\begin{figure}
\includegraphics[width=\columnwidth]{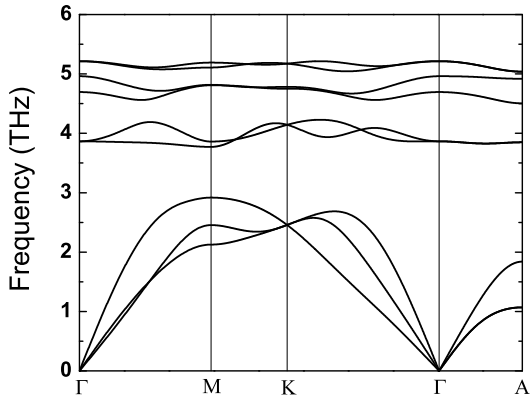}
\caption{Calculated phonon dispersions of IrTe$_2$ in the
HT trigonal structure.}
\label{phonons}
\end{figure}

We also did electronic structure calculations for the LT
experimental structure and for a calculated structure. The calculated
structure was obtained by taking the LT lattice parameters
from experiment and fully relaxing the atomic coordinates to minimize
the total energy. The relaxation was done using the LAPW method in a
scalar relativistic approximation. This structure differed from the
experimental structure in that it has less variation of the Ir bond
valence sums than the experimental structure. However, as seen in the
calculated electronic density of states (Fig. \ref{dos-comp}),
the results
for this structure and the experimental structure are very similar, although
the energy for the calculated structure is lower by 
94 meV per formula unit as compared to the experimental structure.

The DOS shows a strong reduction at $E_F$, from 1.82 eV$^{-1}$ per
formula unit
for the HT structure to 1.00 eV$^{-1}$ per
formula unit for the
experimental LT structure and 1.02 eV$^{-1}$ per
formula unit for
the calculated structure. This sizable change is
consistent with the large changes
in transport and optical properties as observed \cite{PhotoSpecIrTe}
through the phase transition, which is generically what is
expected for a CDW. However, one observes that the DOS
is reconstructed by the
structural distortion
over a large energy range $\sim$ -1.0 eV -- +1.5 eV
(we give energies relative to $E_F$).
This is in contrast to a standard CDW
where the reconstruction is over a range of a few $kT_N$, even
in a strong coupling case.
Furthermore the change in the DOS is not well described as a 
gapping around $E_F$ with a shift of spectral weight to energies
just above or below the gap as in a CDW. Instead there is a downward
shift of all the occupied valence states and a shift of a rather
broad pseudogap in the DOS from the region around $\sim$ 1 eV
to the region around $\sim$ 0 eV.
Furthermore, we do not find a redistribution of the
Ir-Te hybridization. Specifically, the DOS in the gapped region retains
a similar mixture of Ir $5d$ and Te $5p$ character to the
undistorted region. This shows that the distortion is associated
mainly with the Te $5p$ states, rather than Ir-Te bonding.
The large energy range over which the electronic structure is reconstructed
is consistent with recent optical measurements. \cite{PhotoSpecIrTe}
It will be of interest to perform additional spectroscopic measurements,
e.g. by scanning probes and or photoemission,
to compare in detail with the rearrangements of the electronic
structure predicted based on the present crystal structure determinations.
This is characteristic of a rearrangement
of the local bonding, and not a CDW. 

The calculated phonon dispersions of IrTe$_2$ in the HT trigonal
phase are shown in Fig. \ref{phonons}.
As may be seen no unstable branches nor any
unusual dips would suggest Kohn anomalies.
This contradicts what would be expected for a material that undergoes
a CDW.

In conclusion,  the LT structure of IrTe$_2$ was solved by using X-ray/neutron diffraction. The superlattice peaks were observed with the wavevector of (1/5, 0, 1/5) and the associated structural modulation was identified. The first principles calculations based on the LT structure revealed the first
order structural transition in IrTe$_2$ is in a class that strongly
couples to the electronic structure around $E_F$, but is not
related to a Fermi surface instability as in a CDW. Instead it is 
associated with a local bonding instability associated with the
Te $5p$ states. This is consistent with the conclusions of
Fang and co-workers. \cite{PhotoSpecIrTe}

\begin{acknowledgments}
This work was supported by the US Department of Energy, Office of Basic Energy Sciences. HBC and BBC are supported by the Scientific
User Facilities Division,  XC, JQY, MAM, DJS and DGM are supported by the Materials Science and Engineering Division, RC is supported by the Division of
Chemical Sciences, Geosciences, and Biosciences. HY would thank China Scholarship Council for financial assistance.
\end{acknowledgments}

\end{document}